\documentclass[10pt]{article}
\usepackage{fullpage}
\usepackage{amsmath}
\usepackage{amssymb}
\usepackage{amsfonts}
\usepackage{comment}
\usepackage{color}
\usepackage{cite}
\usepackage{subcaption}
\usepackage{graphicx}
\usepackage{enumitem}

\def\blue{\textcolor{blue}}

\def\##1{\underline{#1}}
\def\=#1{\underline{\underline{#1}}}

\def\+
#1{\underline{\bf #1}}
\def\*#1{\underline{\underline{\bf #1}}}

\def\r#1{(\ref{#1})}
\def\l#1{\label{#1}}
\def\c#1{\cite{#1}}

\def\le{\left(}
\def\ri{\right)}
\def\les{\left[}
\def\ris{\right]}
\def\lec{\left\{}
\def\ric{\right\}}
\def\lek{[{\kern 0.1em}}
\def\rik{{\kern 0.1em}]}

\def\.{\mbox{ \tiny{$^\bullet$} }}

\def\eps{\varepsilon}

\def\epso{\eps_{\scriptscriptstyle 0}}
\def\lambdao{\lambda_{\scriptscriptstyle 0}}
\def\muo{\mu_{\scriptscriptstyle 0}}
\def\etao{\eta_{\scriptscriptstyle 0}}

\def\ko{k_{\scriptscriptstyle 0}}

\def\ux{\hat{\#u}_{\rm x}}
\def\uy{\hat{\#u}_{\rm y}}
\def\uz{\hat{\#u}_{\rm z}}

\def\uprop{\hat{\#u}_{\rm prop}}
\def\us{\hat{\#u}_{\rm s}}

\def\calA{{\cal A}}
\def\calB{{\cal B}}
\def\calC{{\cal C}}

\def\PAmat{\lek\=P_\calA\rik}
\def\PBmat{\lek\=P_\calB\rik}
\def\PCmat{\lek\=P_\calC\rik}
\def\Pellmat{\lek\=P_\ell\rik}
\def\Pmat{\lek\=P\rik}
\def\Ymat{\lek \=Y\rik}

\def\fz{\lek\#f(z)\rik}

\def\epsB{\eps_{\calB}}
\def\epsC{\eps_{\calC}}

\def\Req{{\rm Re}\lec{q}\ric}
\def\Imq{{\rm Im}\lec{q}\ric}

\def\dblq#1{\textquotedblleft #1\textquotedblright}

\begin{document}

\begin{center}

\LARGE{ {\bf  Exceptional compound plasmon-polariton   waves 
}}
\end{center}
\begin{center}
\vspace{10mm} \large

  \vspace{3mm}
 {Akhlesh  Lakhtakia}\\
 {\em NanoMM~---~Nanoengineered Metamaterials Group\\ Department of Engineering Science and Mechanics\\
Pennsylvania State University, University Park, PA 16802--6812, USA} 
 \vspace{3mm}\\
 {Chenzhang Zhou}\\
 {\em NanoMM~---~Nanoengineered Metamaterials Group\\ Department of Engineering Science and Mechanics\\
Pennsylvania State University, University Park, PA 16802--6812, USA}
 \vspace{3mm}\\
 {Tom G. Mackay}\footnote{E--mail: T.Mackay@ed.ac.uk.}\\
{\em School of Mathematics and
   Maxwell Institute for Mathematical Sciences\\
University of Edinburgh, Edinburgh EH9 3FD, UK}\\
and\\
 {\em NanoMM~---~Nanoengineered Metamaterials Group\\ Department of Engineering Science and Mechanics\\
Pennsylvania State University, University Park, PA 16802--6812,
USA}

\normalsize

\end{center}

\begin{center}
\vspace{5mm} {\bf Abstract}
\end{center}

 Ordinarily,  a trimaterial structure comprising a sufficiently  thin metal film interposed between
two homogeneous dielectric materials guides
compound plasmon-polariton (CPP) waves,
for which the fields on both sides
of the metal film decay exponentially
with distance from the nearest metal/dielectric interface.
However, if one of the dielectric materials is anisotropic then the trimaterial structure 
can guide an exceptional CPP wave for a particular propagation direction. 
On the
side of the metal film occupied by the anisotropic dielectric material, 
the fields of the exceptional CPP wave decay as the product of a linear function and
an exponential function of the   distance from the nearest metal/dielectric interface.
The canonical boundary-value problem for CPP-wave propagation has been analyzed and solved  numerically; thereby,   the spatial field profiles for exceptional CPP waves 
for a uniaxial-dielectric/metal/isotropic-dielectric   structure  have been established.
\vspace{5mm}

\section{Introduction}
The planar interface of a homogeneous metal and a homogeneous dielectric
material (even air) can guide  surface-plasmon-polariton (SPP) waves at a frequency
at which the real parts of the relative permittivities of the two partnering materials differ in sign \cite{Pitarke}.
In the canonical treatment of the boundary-value problem, each partnering material
occupies a half space. The electromagnetic fields of an SPP wave  drop off exponentially
with distance from the interface. Since the skin depth \cite{Iskander} of a metal is very small,
the metallic half space can be replaced by a sufficiently thick metal film \cite{Turbadar1959,Turbadar1964}.
This replacement~---~which creates a dielectric/metal/dielectric 
trimaterial structure~---~allows the exploitation of SPP waves for optical sensing \cite{Homola},
communication \cite{Sarid,Quail}, and microscopy \cite{Stabler,Somekh}.

Since a metal film has two faces, the waveguiding phenomenon is not as straightforward
as that due to a single metal/dielectric interface \c{Sarid,Jen2011,Akimov2018}. If the metal film is sufficiently thick,
the two metal/dielectric interfaces will not  interact and each could guide an SPP wave all by itself. But, when the metal film is  thin, the two metal/dielectric interfaces will interact to engender  compound plasmon-polariton   (CPP) waves.

The spatial profile of the electric and magnetic fields of a CPP   wave  depends on the constitutive properties of  both dielectric materials  as well as the metal which is assumed to be isotropic. The fields on either side of the metal film   obey the 4$\times$4-matrix ordinary differential equation \c{Berreman,TMMEO}
\begin{equation}
\label{eq1}
\frac{d}{dz}\fz=  
i \Pmat\.\fz\,,
\end{equation}
where $\fz$ is column 4-vector, $\Pmat$ is a 4$\times$4 matrix, $i=\sqrt{-1}$, and the $z$ axis is aligned normal to the metal film. If the dielectric material on a specific side of the metal film is isotropic, then the fields of the CPP wave on that side   decay
exponentially with distance $\vert{z}\vert$ from the metal film. This is because the matrix $\Pmat$ for the dielectric material on that side of the metal film
is semisimply degenerate \cite{Pease}, i.e., it has two distinct eigenvalues, each with algebraic multiplicity equal to two and geometric multiplicity also equal to two.

If the dielectric material on a specific side   of the metal film is anisotropic, then two possibilities arise as follows \cite{Pease}:
\begin{itemize}
\item [I.] The matrix $\Pmat$ for the dielectric material on that side of the metal film
is non-degenerate, i.e., it has four distinct eigenvalues, each with algebraic multiplicity equal to one and geometric multiplicity also equal to one. Then, the fields   on that side   decay
exponentially with distance $\vert{z}\vert$ from the metal film \cite{ESW_book}.

\item[II.] The matrix $\Pmat$ for the dielectric material on that side of the metal film
is non-semisimply degenerate, i.e., it has two distinct eigenvalues, each with algebraic multiplicity equal to two but geometric multiplicity equal to one. Then, the fields  on that side   
vary as the products of a linear function and an exponential function
of the distance $\vert{z}\vert$ from the metal film \cite{LM2020},   decaying as $\vert{z}\vert \to\infty$  \cite{ESW_book}.

\end{itemize}
Case~I is commonplace, but this paper introduces Case~II for CPP-wave propagation guided by a metal film interposed between two homogeneous dielectric materials.  As
non-semisimple degeneracy cannot be exhibited by an isotropic dielectric material, at least one of the two dielectric materials must be anisotropic. There are no other restrictions on that anisotropic material: it can be dissipative, active, or neither dissipative nor active. 

In this paper, we consider CPP-wave propagation when one of the two dielectric materials
(labeled $\calA$) is uniaxial with its optic axis aligned normal to the thickness direction of the metal film
and the other dielectric material (labeled $\calC$) is isotropic,  the two being separated by a film of a metal (labeled $\calB$). The matrixes $\PBmat$ and $\PCmat$ for materials $\calB$ and $\calC$, respectively, are semisimply degenerate. When the matrix $\PAmat$ is  non-semisimply degenerate, the CPP wave may be classified as \textit{exceptional},
following the terminology used first
in condensed-matter physics \cite{Moiseyev,Kawabata} and now increasingly in classical electromagnetics \cite{Kiselev2014,Grundmann2017,Hanson2019,LM2020}. When the matrix $\PAmat$ is  either non-degenerate or semisimply degenerate, the CPP wave is
\textit{unexceptional}.

Theory is presented in Sec.~\ref{S2} and illustrative numerical results are provided and discussed in Sec.~\ref{S3}. The paper closes with some remarks in Sec.~\ref{S4}. Throughout the paper,
 the   free-space permittivity, permeability,
 wavenumber, wavelength,  and impedance are written as $\epso$, $\muo$,
 $\ko = \omega \sqrt{\epso \muo}$, $\lambdao = 2 \pi / \ko$, and $\etao = \sqrt{\muo/\epso}$, respectively, with 
   $\omega$ being  the angular frequency. 
Single underlining 
with no enclosing square brackets
signifies a 3-vector. The position vector
is denoted by $\#r=x\ux+y\uy+z\uz$, where
 $\lec \ux, \uy, \uz \ric$ is
the triad of unit vectors aligned with the Cartesian axes. 
 Double underlining
with no enclosing square brackets
signifies a 3$\times$3 dyadic \cite{Chen}.
Matrixes and column vectors are
double and single underlined, respectively, and
 enclosed by square brackets. 
The superscript ${}^T$ denotes the transpose.
The  operators $\mbox{Re} \lec \. \ric$ 
and $\mbox{Im} \lec \. \ric$ deliver the
real and imaginary parts, respectively, of  complex-valued quantities;  
 the complex conjugate is denoted by an asterisk;   and
  dependence on time $t$ is achieved implicitly through
   $\exp(-i\omega t)$.

\section{Theory} \l{S2}

\subsection{ 4$\times$4 matrix ordinary-differential-equation formalism}

\begin{figure}[!htb]
\centering

 \includegraphics[width=7.3cm]{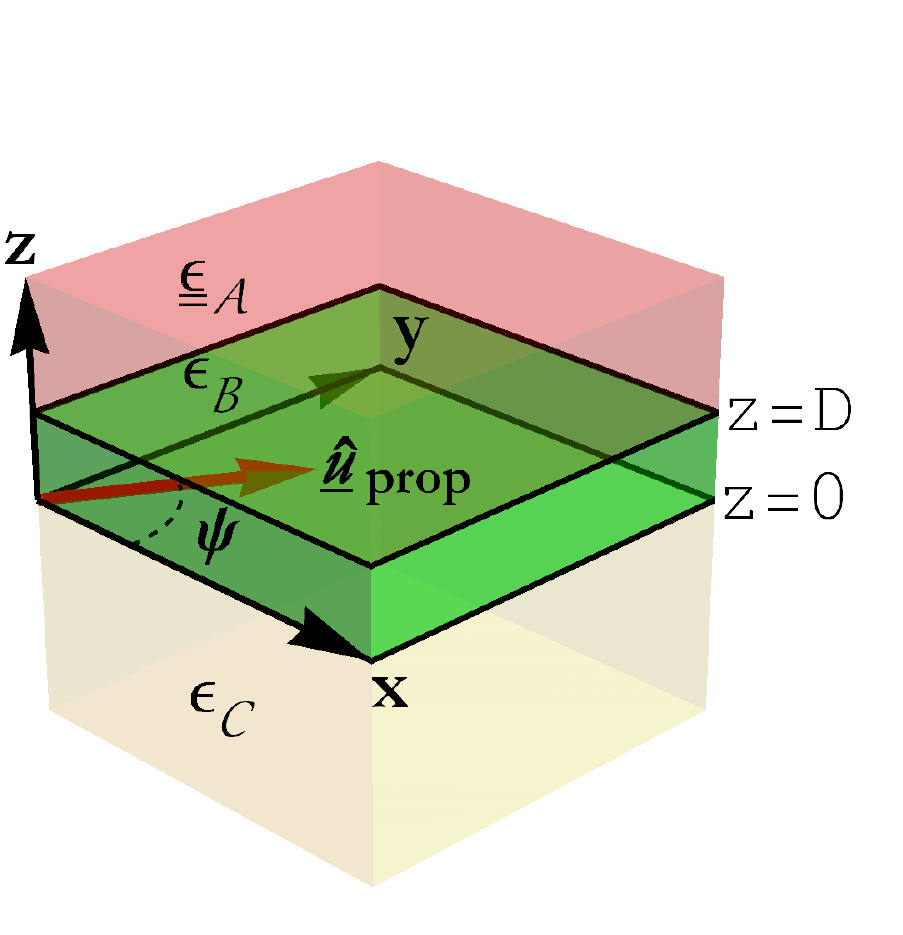} 

 \caption{\label{Schematic_diagram} 
Schematic  of the canonical boundary-value problem for 
the propagation of CPP waves   parallel to the unit vector $\uprop$ that lies
wholly in the $xy$ plane at an angle $\psi$ relative to the $x$ axis.
}
\end{figure}

The dielectric material $\calA$  fills the half-space $z>D$, the dielectric material $\calC$   fills
the half-space $z<0$, the two being separated by the metal $\calB$ in the region $0<z<D$,
as shown in Fig.~\ref{Schematic_diagram}.

Material $\calA$ is  uniaxial dielectric   specified by the relative permittivity dyadic \cite{Chen}
\begin{equation}
\label{epsA}
\=\eps_\mathcal{A} = \eps_\mathcal{A}^{\rm s} \=I + \le
\eps_\mathcal{A}^{\rm t} - \eps_\mathcal{A}^{\rm s} \ri \,
\ux \, \ux
\,,
 \end{equation}
 with 
 $\eps_\mathcal{A}^{\rm s}\in\mathbb{C}$ and $\eps_\mathcal{A}^{\rm t}\in\mathbb{C}$ being 
 the principal relative permittivity scalars and $\=I$ being the identity 3$\times$3 dyadic.   We set $\mbox{Re} \lec  \eps_\mathcal{A}^{\rm s} \ric>1$,
 $\mbox{Re} \lec  \eps_\mathcal{A}^{\rm t} \ric>1$,
 $\mbox{Im} \lec   \eps_\mathcal{A}^{\rm s} \ric\geq0$,
 and $\mbox{Im} \lec    \eps_\mathcal{A}^{\rm t}  \ric\geq0$.
 The intermediate material $\calB$ is a metal with relative permittivity $\eps_\calB\in\mathbb{C}$
such that $\mbox{Re} \lec \eps_\calB \ric<0$ and $\mbox{Im} \lec \eps_\calB \ric>0$. Finally, the   dielectric
material $\calC$ is   isotropic and is  characterized by the relative permittivity $\eps_\calC\in\mathbb{C}$ with
 $\mbox{Re} \lec \eps_\calC \ric>1$ and $\mbox{Im} \lec \eps_\calC \ric\geq0$.


The electric and magnetic field phasors for CPP-wave propagation 
are expressed everywhere as
\cite{ESW_book} 
\begin{equation} \label{planewave}
\left.\begin{array}{l}
 \#E (\#r)=  \les e_x(z)\,\ux + e_y(z)\,\uy+e_z(z)\,\uz \ris \,  
  \exp\left( i q\, \uprop\.\#r\right) \\[4pt]
 \#H  (\#r)=  \les h_x(z)\,\ux + h_y(z)\,\uy+h_z(z)\,\uz \ris \,
  \exp\left( i q\, \uprop\.\#r\right)
 \end{array}\right\}\,,  \:\:\:   z\in(-\infty,\infty)\,,
\end{equation}
with ${q}$ being the guide  wavenumber. Relative to  the $x$ axis, 
the direction of propagation in the  $xy$ plane is identified
by the unit vector
\begin{equation}
\uprop=\ux\cos\psi+\uy\sin\psi\,,
\end{equation}
 where the
 angle $\psi\in\left[0,2\pi\right)$.
Substitution of  the phasor representations~\r{planewave} in
the source-free Maxwell curl equations
yields
 the 4$\times$4 matrix ordinary differential
equations \c{Berreman,TMMEO}
\begin{equation}
\label{MODE_A}
\frac{d}{dz}\fz= \left\{
\begin{array}{l}
i \PAmat\.\fz\,,  \qquad   z>D 
\vspace{8pt} \\
i \PBmat\.\fz\,,  \qquad   0<z<D
\vspace{8pt} \\
i \PCmat\.\fz\,,  \qquad   z<0
\end{array} , \right.
\end{equation}
wherein
the  column 4-vector
\begin{equation}
\fz= 
\lek
\begin{array}{c}%
e_ x(z), \quad
e_y(z),\quad
h_x(z),\quad
h_y(z)
\end{array}
\rik^T,
\label{f-def}
\end{equation}
and  the 4$\times$4 propagation matrixes  \cite{MCL2019}
\begin{equation}
\hspace{-0.5cm}
\small{\PAmat=  \les   
\begin{array}{cccc}
0&0& \displaystyle{ \frac{q^2 \cos \psi \sin \psi}{\omega \epso \eps^s_\calA}} & 
\displaystyle{\frac{\ko^2 \eps^s_\calA- q^2 \cos^2 \psi }{\omega \epso \eps^s_\calA} } \vspace{8pt} \\
0&0& \displaystyle{\frac{-\ko^2 \eps^s_\calA+ q^2 \sin^2 \psi }{\omega \epso \eps^s_\calA} }&
\displaystyle{ -\frac{q^2 \cos \psi \sin \psi}{\omega \epso \eps^s_\calA}} \vspace{8pt}
\\
\displaystyle{ -\frac{q^2 \cos \psi \sin \psi}{\omega \muo}} & 
\displaystyle{\frac{-\ko^2 \eps^s_\calA+ q^2 \cos^2 \psi }{\omega \muo} } &0&0\vspace{8pt} \\
\displaystyle{\frac{\ko^2 \eps^t_\calA- q^2 \sin^2 \psi }{\omega \muo} } &
\displaystyle{ \frac{q^2 \cos \psi \sin \psi}{\omega \muo}}&0&0
\end{array}
\ris}
\end{equation}
and
\begin{eqnarray}  
\nonumber
&& \hspace{-0.95cm}
\small{\Pellmat=
\les   
\begin{array}{cccc}
0&0& \displaystyle{ \frac{q^2 \cos \psi \sin \psi}{\omega \epso \eps_\ell}} & 
\displaystyle{\frac{\ko^2 \eps_\ell- q^2 \cos^2 \psi }{\omega \epso \eps_\ell} } \vspace{8pt} \\
0&0& \displaystyle{\frac{-\ko^2 \eps_\ell+ q^2 \sin^2 \psi }{\omega \epso \eps_\ell} }&
\displaystyle{ -\frac{q^2 \cos \psi \sin \psi}{\omega \epso \eps_\ell}} \vspace{8pt}
\\
\displaystyle{ -\frac{q^2 \cos \psi \sin \psi}{\omega \muo}} & 
\displaystyle{\frac{-\ko^2 \eps_\ell+ q^2 \cos^2 \psi }{\omega \muo} } &0&0\vspace{8pt} \\
\displaystyle{\frac{\ko^2 \eps_\ell- q^2 \sin^2 \psi }{\omega \muo} } &
\displaystyle{ \frac{q^2 \cos \psi \sin \psi}{\omega \muo}}&0&0
\end{array}
\ris},
\\[5pt]
&& \hspace{8cm}
\quad \ell\in\lec\calB,\calC\ric\,.
\end{eqnarray}
Whereas
\begin{equation}
h_ z(z) = \displaystyle{\frac{q \les e_ y(z) \cos \psi - e_ x(z)  \sin \psi  \ris}{\omega \muo }}\,,\qquad
z\in(-\infty,\infty)\,,
\end{equation}
holds in all three regions,
\begin{equation}
e_z(z) =
\left\{
\begin{array}{ll}
 \displaystyle{-\,\frac{q \les h_ y(z) \cos \psi - h_ x(z)  \sin \psi  \ris}{\omega \epso \eps^s_\calA}}\,,\qquad z > D\,,
  \vspace{8pt} \\
   \displaystyle{-\,\frac{q \les h_ y(z) \cos \psi - h_ x(z)  \sin \psi  \ris}{\omega \epso \epsB}}\,,\qquad 0<z < D\,,
  \vspace{8pt} \\
 \displaystyle{-\,\frac{q \les h_ y(z) \cos \psi - h_ x(z)  \sin \psi  \ris}{\omega \epso \epsC}}\,,\qquad z <0\,.
\end{array}
\right. 
\end{equation}

\subsection{Fields in material $\calA$}
The four eigenvalues of  $\PAmat$ can be written as $\pm \alpha_{\calA 1}$ and $\pm \alpha_{\calA 2}$. 
The two with positive imaginary parts are  
\begin{equation} \l{a_decay_const}
\left.
\begin{array}{l}
\alpha_{\calA 1} = i \sqrt{ q^2 - \ko^2 \eps_\calA^s} \vspace{8pt}\\
\alpha_{\calA 2} = \displaystyle{
i\sqrt{\frac{q^2 \les \le \eps^s_\calA + \eps^t_\calA \ri  - \le \eps^s_\calA - \eps^t_\calA \ri \cos 2 \psi \ris   - 2 \ko^2 \eps^s_\calA \eps^t_\calA}{2 \eps^s_\calA}}
}
\end{array}
\right\}\,,
\end{equation}
When $\alpha_{\calA 1}\ne \alpha_{\calA 2}$, the column vectors
\begin{equation}
\lek\#v_{\calA 1}\rik = 
\les \begin{array}{c}
0 \vspace{8pt}\\
\displaystyle{
\frac{\ko \alpha_{\calA 1}}{q^2 \sin \psi \cos \psi}} \vspace{8pt}\\
\displaystyle{\frac{\cot 2 \psi}{\etao} + \frac{\csc 2 \psi}{\etao}  \le 1 - \frac{2 \ko^2 \eps^s_\calA }{q^2} \ri}\vspace{8pt} \\ \etao^{-1}
\end{array}
\ris
\end{equation}
and
\begin{equation}
\lek\#v_{\calA 2}\rik = 
\les \begin{array}{c}
\displaystyle{1 - \frac{q^2 \le \cos 2 \psi + 1 \ri}{2\ko^2
 \eps^s_\calA 
}} \vspace{8pt}\\
\displaystyle{-\frac{q^2  \cos  \psi \sin \psi }{ \ko^2 \eps^s_\calA }} \vspace{8pt}\\ 0 \vspace{8pt} \\ 
\displaystyle{\frac{\alpha_{\calA 2}}{\omega \muo}}
\end{array}
\ris
\end{equation}
are the eigenvectors of $\PAmat$   matching the eigenvalues $+\alpha_{\calA 1}$ and $+\alpha_{\calA 2}$,
respectively,
Hence, the general solution to Eq.~\r{MODE_A}${}_1$   is given as \cite{ESW_book}
\begin{equation} 
\l{D_gen_sol}
\fz =C_{\calA 1} \lek \#v_{\calA 1} \rik\exp \les i \alpha_{\calA 1} (z-D) \ris +  
C_{\calA 2}\lek \#v_{\calA 2}\rik \exp \les i \alpha_{\calA 2} (z-D) \ris \,,\qquad z > D\,,
\end{equation}
for   fields that decay as $z \to +\infty$. 
The complex-valued constants $C_{\calA 1}$ and $C_{\calA 2}$ have to be determined by 
application of appropriate boundary conditions at the plane $z=D$.

When $\PAmat$ exhibits
non-semisimple degeneracy,
\begin{equation}
\l{alphaa_sol} 
\alpha_{\calA 1} = \alpha_{\calA 2} \equiv \alpha_\calA = iq\, \sin \psi
\end{equation}
and
\begin{equation} 
\l{q_sol}
q = {\rm sgn}(\cos\psi) \frac{\ko  \sqrt{ \eps^s_{\calA}}}{\cos \psi}\,,
 \end{equation} 
where ${\rm sgn}(\zeta) = 1$ if $\zeta>0$ and   ${\rm sgn}(\zeta) =- 1$ if $\zeta<0$.
The square root in Eq.~\r{q_sol} must be chosen
to ensure that $\mbox{Im} \lec \alpha_\calA \ric > 0$.  
The general solution of Eq.~\r{MODE_A}${}_1$ is then expressed as \cite{MCL2019}
\begin{equation} \l{DV_gen_sol}
\fz=  \left( C_{\calA 1}  \lek\#v _{\calA }\rik  +  \ko\, C_{\calA 2}
\lec i  (z-D) \, \lek\#v _{\calA} \rik  + \lek\#w_{\calA}\rik \ric \right)
 \exp \les i \alpha _{\calA } (z-D) \ris \,,\quad z > D\,,
\end{equation}
for fields that decay as $z \to +\infty$,
where
\begin{equation} 
\l{vA}
\lek \#v_{\calA } \rik =  
\les \begin{array}{c}
0 \vspace{8pt}\\
\displaystyle{{\rm sgn}(\cos\psi)\frac{ i }{\sqrt{ \eps^s_\calA}}} \vspace{8pt}\\
0 \vspace{8pt} \\ \etao^{-1}
\end{array}
\ris
\end{equation}
and
\begin{equation} 
\l{wA}
  \lek \#w_{\calA } \rik  = \frac{1}{\ko}
\les \begin{array}{c}
\displaystyle{  \frac{2}{ \eps^t_\calA - \eps^s_\calA  }}
\vspace{8pt}\\
\displaystyle{\frac{\tan \psi}{ \eps^s_\calA} \le \cot^2 \psi - 2\frac{ \eps_\calA^s - \eps^t_\calA \cot^2 \psi }{\eps_\calA^s - \eps^t_\calA}
\ri
} \vspace{8pt}\\
\displaystyle{ {\rm sgn}(\cos\psi) \frac{2 i \sqrt{\eps^s_\calA}}{\etao \le \eps^t_\calA - \eps^s_\calA \ri}}
 \vspace{8pt} \\ 0
\end{array}
\ris.
\end{equation}

\subsection{ Fields in material $\calB$}

The  4$\times$4 matrix $\PBmat$    
cannot exhibit non-semisimple degeneracy, and  the general solution of Eq.~\r{MODE_A}${}_2$ 
can be stated as \cite{TMMEO}
\begin{equation}
\l{eq16}
\fz = \exp\lec i \PBmat z\ric\.\lek \#f(0^+)\rik\,, \quad 0 < z < D\,,
\end{equation}
which yields
\begin{equation}
\l{eq17}
\lek \#f(D^-)\rik = \exp\lec i \PBmat D\ric\.\lek \#f(0^+)\rik\,.
\end{equation}

\subsection{ Fields in material $\calC$}

The  4$\times$4 matrix $\PCmat$    has two distinct eigenvalues $\pm \alpha_{\calC}$, where
\begin{equation} \l{b_decay_const}
\alpha_{\calC} =- i \sqrt{q^2 - \ko^2 \eps_\calC} .
\end{equation}
 The sign of the square root  in Eq.~\r{b_decay_const} must be such that 
 $\mbox{Im} \lec \alpha_{\calC} \ric < 0 $ for CPP-wave propagation.
The two linearly independent  eigenvectors of 
$\PCmat$  corresponding to the eigenvalue $\alpha_{\calC}$ 
are given by
\begin{equation}
\left.
\begin{array}{l}
 \lek\#v_{\calC 1}\rik = 
\les 
\displaystyle{1 - \frac{  q^2 \cos^2 \psi}{\ko^2 \eps_\calC}}, \quad
\displaystyle{- \frac{  q^2 \cos \psi \sin \psi}{\ko^2 \eps_\calC}}, \quad 0, \quad
 \displaystyle{\frac{\alpha_\calC}{\omega \muo}}
\ris^T
\vspace{8pt}
\\
 \lek\#v_{\calC 2}\rik = 
\les 
\displaystyle{ \frac{  q^2 \cos \psi \sin \psi}{\ko^2 \eps_\calC}}, \quad
\displaystyle{ \frac{ q^2 \sin^2 \psi}{\ko^2 \eps_\calC} - 1}, \quad
 \displaystyle{\frac{\alpha_\calC}{\omega \muo}, \quad 0}\:\:
\ris^T
\end{array}
\right\}.
\end{equation}
Hence,   
\begin{equation}
\label{2.22-AL}
\fz = \lec C_{\calC 1} \lek \#v_{\calC 1}  \rik +  C_{\calC 2} \lek\#v_{\calC 2}\rik \ric \exp \le i \alpha_{\calC} z \ri 
\,,\qquad z < 0\,,
\end{equation}
is  the general solution of Eq.~\r{MODE_A}${}_3$ for fields 
  that decay as $z \to -\infty$, wherein
 the complex-valued constants $C_{\calC 1}$ and $C_{\calC 2}$ have to be determined by applying boundary conditions at 
 the plane $z=0$.

\subsection{ Boundary conditions}

The tangential  components of the electric and magnetic field
 phasors must be continuous across the interface planes $z=0$ and $z=D$; i.e.,
 \begin{equation}
 \label{eq21}
 \lek\#f(0^+)\rik=  \lek\#f(0^-)\rik
 \end{equation} 
and
 \begin{equation}
 \label{eq22}
 \lek\#f(D^+)\rik=  \lek\#f(D^-)\rik
 \,.
 \end{equation} 
 The use of Eq.~\r{eq17} therefore yields \cite{ESW_book}
\begin{equation}
 \label{eq23}
 \lek\#f(D^+)\rik=  \exp\lec i \PBmat D\ric\.\lek \#f(0^-)\rik
 \,,
 \end{equation} 
 where
 \begin{equation}
 \lek \#f(0^-)\rik=
 C_{\calC 1} \lek \#v_{\calC 1}  \rik +  C_{\calC 2} \lek\#v_{\calC 2}\rik   
\end{equation}
by virtue of 
Eq.~\r{2.22-AL}. Next, either
\begin{equation} 
\l{D_gen_sol-1}
 \lek\#f(D^+)\rik =C_{\calA 1} \lek \#v_{\calA 1} \rik  +  
C_{\calA 2}\lek \#v_{\calA 2}\rik   
\end{equation}
from Eq.~\r{D_gen_sol} leading to
\begin{equation} 
\l{D_gen_sol-2}
C_{\calA 1} \lek \#v_{\calA 1} \rik  +  
C_{\calA 2}\lek \#v_{\calA 2}\rik =
 \exp\lec i \PBmat D\ric\. \lec C_{\calC 1} \lek \#v_{\calC 1}  \rik +  C_{\calC 2} \lek\#v_{\calC 2}\rik \ric
\end{equation}
or
\begin{equation} 
\l{DV_gen_sol-1}
 \lek\#f(D^+)\rik=  C_{\calA 1}  \lek\#v_{\calA }\rik  +  \ko\, C_{\calA 2}
  \lek\#w_{\calA}\rik 
\end{equation}
from Eq.~\r{DV_gen_sol} delivering
\begin{equation}
\l{DV_gen_sol-2}
C_{\calA 1}  \lek\#v_{\calA }\rik   +  \ko\, C_{\calA 2}
   \lek\#w_{\calA}\rik   =
 \exp\lec i \PBmat D\ric\. \lec C_{\calC 1} \lek \#v_{\calC 1}  \rik +  C_{\calC 2} \lek\#v_{\calC 2}\rik \ric\,.
\end{equation}

Both Eq.~\r{D_gen_sol-2} and Eq.~\r{DV_gen_sol-2} can be put in the form
\begin{equation}
\Ymat\. \les \:
 C_{\calA 1}, \quad
  C_{\calA 2}, \quad
   C_{\calC 1}, \quad
    C_{\calC 2} \:
 \ris^T =  \les \:
 0, \quad
  0, \quad
   0, \quad
    0 \:
 \ris^T.
\end{equation}
As the 4$\times$4 characteristic matrix  $\Ymat$   must be singular for  CPP-wave propagation,
the dispersion equation 
\begin{equation}
\l{dispersion_eq}
\left\vert  \Ymat \right\vert = 0
 \end{equation} 
 emerges.
 
 If $\psi$ is replaced by $-\psi$ or by $\pi \pm \psi$ then 
the dispersion equation~\r{dispersion_eq} is unchanged. Accordingly, in the following 
numerical investigation of unexceptional and exceptional CPP waves, attention is restricted to the quadrant $0 \leq \psi \leq \pi/2$.

\section{Numerical results and discussion} \l{S3}

All calculations were made for $\lambdao=633$~nm fixed.
Silver was chosen as the metal so that
$\epsB = -16.07 + 0.44 i$ \cite{silver}. As the skin depth of silver then is $25.11$~nm \cite{Iskander}, the  thickness $D$ was varied in
the range $(0,80]$~nm. The constitutive parameters 
$\eps^s_\calA$, $\eps^t_\calA$, and $\eps_\calC$ were varied to bring out diverse facets of
the CPP waves under investigation. In particular,   material $\calA$ was chosen to be  an effectively homogeneous    material comprising electrically small spheroidal inclusions  distributed in a host material; by varying the volume fraction and elongation of the inclusions, as well as the permittivities of the inclusion and host materials, the constitutive parameters 
$\eps^s_\calA$ can be  $\eps^t_\calA$ adjusted \cite{Neelakanta,Mackay2011}. In contrast,
   material $\calC$ was chosen to be a natural one.  The angle  $\psi \in[0,\pi/2]$  to fix  
  the direction of propagation was varied to find a value at which an exceptional CPP wave can exist.

\begin{figure}[!htb]
\centering
\includegraphics[width=4cm]{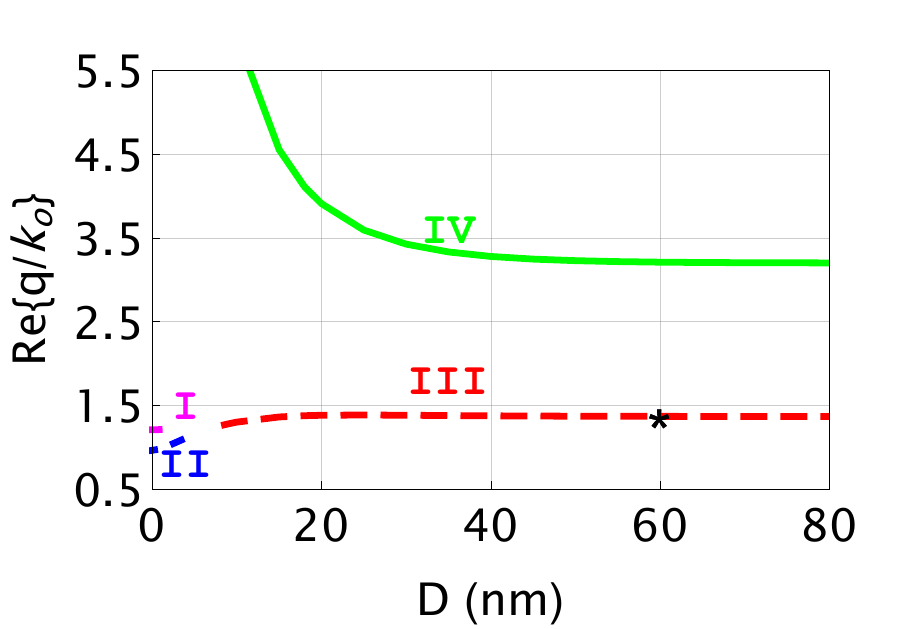} 
\includegraphics[width=4cm]{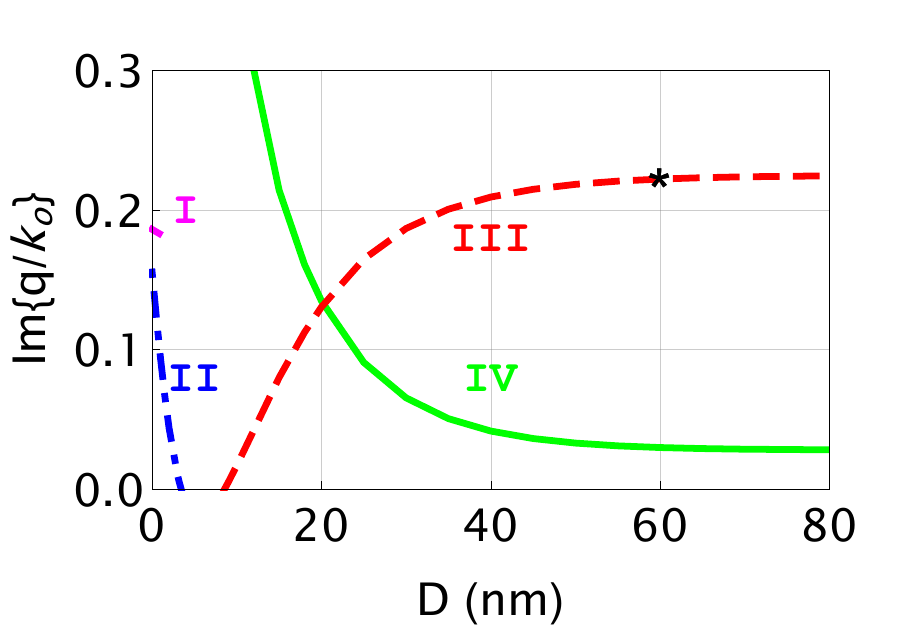}
 \caption{\label{qFigure2} 
${\rm Re}\lec{q/\ko}\ric$ and ${\rm Im}\lec{q/\ko}\ric$ plotted versus
$D\in(0,80]$~nm
for CPP waves  when $\eps^s_\calA=1.5+0.5i$, $\eps^t_\calA=3.1282+0.1111i$, $\epsB = -16.07 + 0.44 i$, $\eps_\calC=6.26$, $\psi=25^\circ$, and $\lambdao=633$~nm.
The solitary exceptional point of $\lek{\underline{\underline P}}_\calA\rik$ is identified by a black star in both plots. The solution branches are numbered I to IV.    }
\end{figure}

Let us begin by choosing  $\eps^s_\calA=1.5+0.5i$, $\eps^t_\calA=3.1282+0.1111i$, and $\eps_\calC=6.26$ (zinc selenide \cite{Marple}). The solutions of Eq.~\r{dispersion_eq} as functions
of  $D\in(0,80]$~nm for $\psi=25^\circ$ are organized
into four branches in Fig.~\ref{qFigure2} as follows:
\begin{itemize}
\item [I.] The shortest branch commences at  $D=0^+$  and
terminates at  $D\simeq0.83$~nm.
\item[II.] The next longer branch commences at $D=0^+$ and terminates at $D\simeq3.43$~nm.
\item[III.] The next longer branch begins at $D\simeq 8.55$~nm. After
$D$ increases beyond $60$~nm, the solution on this branch tends towards
$(1.3679 +0.2251 i)\ko$, which is the wavenumber of the SPP wave guided solely
by the $\calA/\calB$ interface   \cite{ZML_pra}.
\item[IV.] The longest branch commences at  $D=0^+$ and the solution
on this branch tends towards  $q=(3.2012+0.0279i)\ko$ as $D$ increases, which is the wavenumber of the SPP wave guided solely
by the $\calB/\calC$ interface  \cite{Pitarke,ESW_book}.

\end{itemize}
Both (i) the existence of Branches I and II as well as (ii) the deviation of Branches III and IV from their respective asymptotes indicate the interaction of the $\calA/\calB$ and $\calB/\calC$ interfaces in the creation of CPP waves that are not merely the spatial superpositions of the SPP waves guided either by the $\calA/\calB$ interface by itself or the $\calB/\calC$ interface by itself.

Hence, CPP waves can be said to exist for $D>0$. These are of the unexceptional kind, except that the CPP wave on Branch~III for $D = 60$~nm is exceptional because $\PAmat$ exhibits
a non-semisimple degeneracy. The fact that the exceptional CPP wave exists on Branch~III alone was unsurprising in retrospect, because  the solution on this branch
tends towards $q \simeq(1.3679 +0.2251 i)\ko$ and  the  $\calA/\calB$ interface by itself can support the existence of an exceptional SPP wave with  $q = ( 1.3695 + 0.2222 i )\ko$ \cite{ZML_pra}.

The surface wave guided by the interface
of materials $\calA$ and $\calC$ when $D=0$ is classified as a Dyakonov surface wave \cite{Marchevskii,Dyakonov,ESW_book}. The wavenumbers of the two Dyakonov surface waves 
guided by the $\calA/\calC$ interface by itself
are $  q = (1.2095 + 0.1862 i)\ko $   and $q=(0.9642 + 0.1556i )\ko$ \cite{Zapata,Temp}. 
Thus, both Branches I and II of unexceptional CPP waves can be extended to include the Dyakonov surface waves that exist for $D=0$.

Both $\Req$ and $\Imq$ on Branch~IV rise
monotonically and rapidly  as $D\to0^+$. Thus, the phase speed $\omega/{\rm Re}\lec{q}\ric$ decreases and the attenuation rate ${\rm Im}\lec{q}\ric$ increases \cite{ESW_book},
and the unexceptional CPP wave becomes ineffective as a transporter of electromagnetic energy.

The matrix $\PAmat$ exhibits non-semisimple degeneracy 
in Fig.~\ref{qFigure2} at a  value of $D$ that is more than twice the
skin depth of silver, so that the exceptional CPP wave is almost
an exceptional SPP wave guided by the $\calA/\calB$ interface all by itself. In order for
the exceptional CPP wave to occur at a smaller value of $D$ so that the
$\calA/\calB$ and $\calB/\calC$ interfaces interact, we changed
the extraordinary relative permittivity scalar \cite{Chen}
  of material $\calA$ to $\eps^t_\calA=1.6173+ 0.6659 i$
  and chose perfluorohexane (C$_6$F$_{14}$) as material $\calC$ so that
  $\epsC=1.5625$ \cite{Fluorinert}. All other parameters were left the same
  as for Fig.~\ref{qFigure2}.
  
\begin{figure}[!htb]
\centering
\includegraphics[width=4cm]{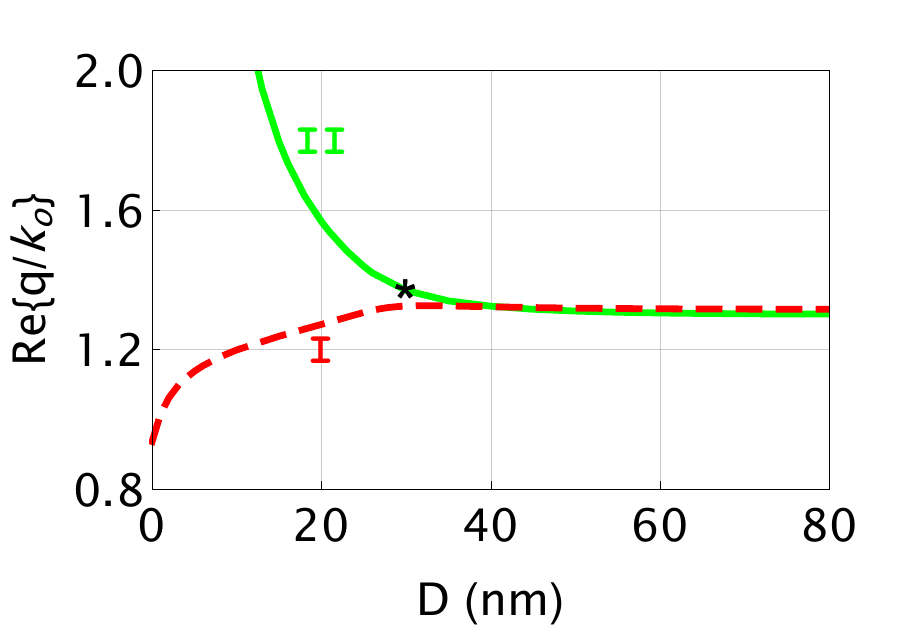} 
\includegraphics[width=4cm]{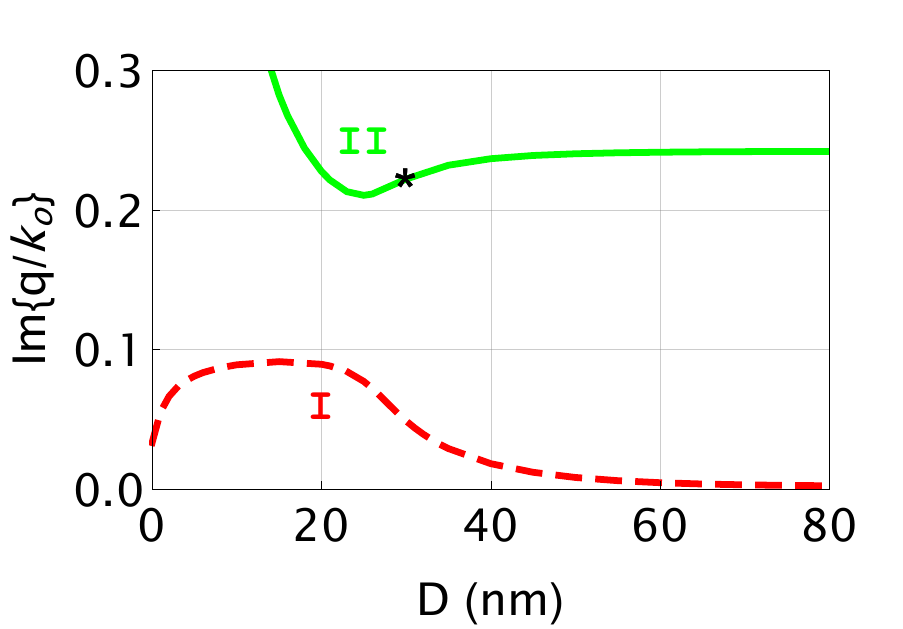}
 \caption{\label{qFigure3} 
${\rm Re}\lec{q/\ko}\ric$ and ${\rm Im}\lec{q/\ko}\ric$ plotted versus
$D\in(0,80]$~nm
for CPP waves  when $\eps^s_\calA=1.5+0.5i$, $\eps^t_\calA=1.6173+ 0.6659 i$, $\epsB = -16.07 + 0.44 i$, $\eps_\calC=1.5625$, $\psi=25^\circ$, and $\lambdao=633$~nm.
The solitary exceptional point of $\lek{\underline{\underline P}}_\calA\rik$ is identified by a black star in both plots. The solution branches are numbered I and II. 
  }  
\end{figure}

The solutions of Eq.~\r{dispersion_eq} as functions
of  $D\in(0,80]$~nm are organized
into two branches in Fig.~\ref{qFigure3} as follows:
\begin{itemize}

\item[I.] This branch   starts at $D=0^+$. Both $\Req$ and $\Imq$ increase as $D$ increases to $\sim20$~nm. As $D$ increases further, $\Req$ increases but $\Imq$ decreases and the solution on this branch tends towards
 $(1.3155 +0.0019 i)\ko$, which is the wavenumber of the SPP wave guided solely
by the $\calB/\calC$ interface  \cite{Pitarke,ESW_book}.

\item[II.] This branch starts at $D=0^+$ with  very large values of $\Req$ and $\Imq$ so that
the corresponding CPP wave is not an effective transporter of electromagnetic energy. However,
both $\Req$ and $\Imq$ decline rapidly and do not change significantly for $D> 50$~nm.
After
$D$ increases beyond $40$~nm, the solution on this branch tends towards
 $(1.3011 +0.2422 i)\ko$, which is the wavenumber of the SPP wave guided solely
by the $\calA/\calB$ interface   \cite{ZML_pra}.
The matrix $\PAmat$ exhibits
a non-semisimple degeneracy at  $D=30$~nm, giving rise to an exceptional CPP
wave on Branch~II with  $q=(1.3695 + 0.2222 i)\ko$. 
\end{itemize}

The planar interface of materials $\calA$ and $\calC$ can guide a Dyakonov surface wave in the direction specified by $\psi=25^\circ$. The wavenumber of this surface wave is
$q=(  0.9364 + 0.0334 i)\ko$, which means that Branch~I  of unexceptional CPP waves can be extended to include the Dyakonov surface wave that exists for $D=0$.

\begin{figure}[!htb]
\centering
 \includegraphics[width=4cm]{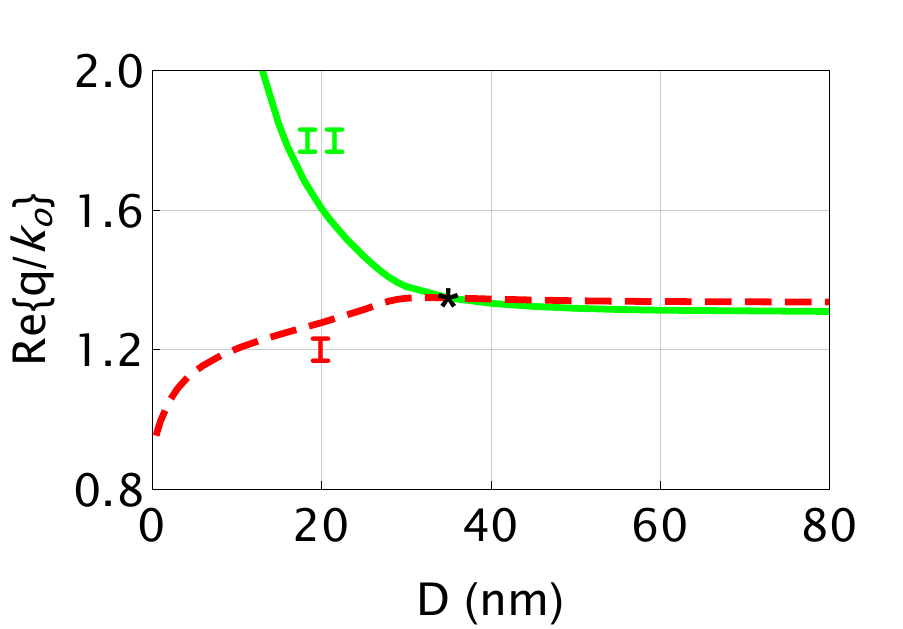} 
 \includegraphics[width=4cm]{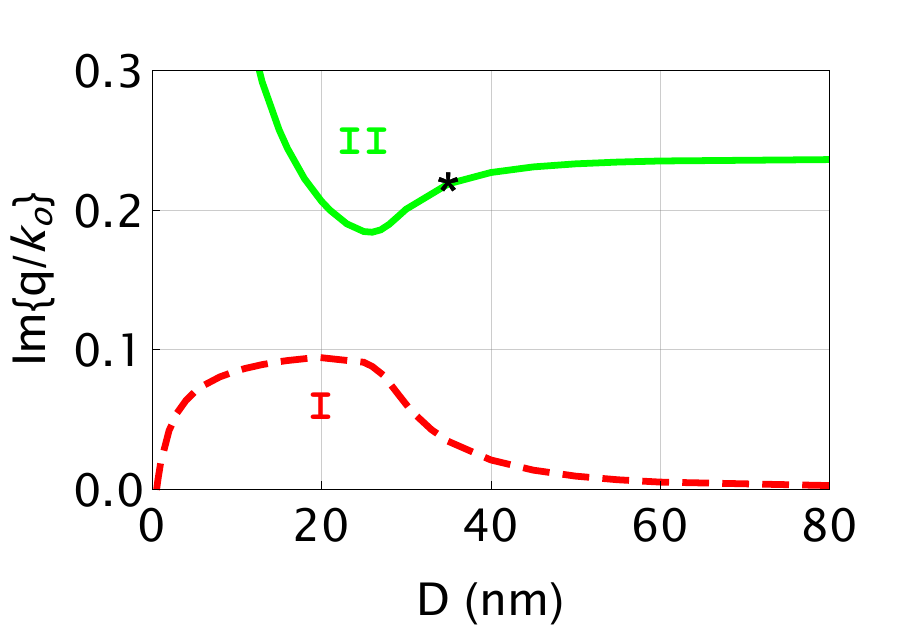}
 \caption{\label{qFigure4} 
${\rm Re}\lec{q/\ko}\ric$ and ${\rm Im}\lec{q/\ko}\ric$ plotted versus 
$D\in(0,80]$~nm
for CPP waves  when $\eps^s_\calA=1.5+0.5i$, $\eps^t_\calA=1.7896 + 0.4807 i$, $\epsB = -16.07 + 0.44 i$, $\eps_\calC=1.6066$, $\psi=23^\circ$, and $\lambdao=633$~nm.
The solitary exceptional point of $\lek{\underline{\underline P}}_\calA\rik$ is identified by a black star in both plots. The solution branches are numbered I and II.
 }
\end{figure}

Finally, we modified the constitutive parameters to $\eps^t_\calA=1.7896+ 0.4807 i$
and $\epsC=1.6066$ and the direction of propagation to $\psi=23^\circ$, in order to obtain
an exceptional CPP wave and an unexceptional CPP wave with identical phase speeds
 for the same value of $D$. The solutions of Eq.~\r{dispersion_eq} as functions
of  $D\in(0,80]$~nm are organized
into two branches in Fig.~\ref{qFigure4} as follows:
\begin{itemize}

\item[I.] This branch   starts at  $D=0.58$~nm. Both $\Req$ and $\Imq$ increase as $D$ increases to $\sim20$~nm. As $D$ increases further, $\Req$ increases but $\Imq$ decreases and the solution on this branch tends towards
 $(1.3360 +0.0020 i)\ko$, which is the wavenumber of the SPP wave guided solely
by the $\calB/\calC$ interface  \cite{Pitarke,ESW_book}.

\item[II.] This branch starts at $D=0^+$ with  very large values of $\Req$ and $\Imq$ so that
the corresponding CPP wave is an ineffective transporter of electromagnetic energy. However,
both $\Req$ and $\Imq$ decline rapidly.
After
$D$ increases beyond $40$~nm, the solution on this branch tends towards
$(1.3089 +0.2366 i)\ko$, which is the wavenumber of the SPP wave guided solely
by the $\calA/\calB$ interface   \cite{ZML_pra}.
\end{itemize}

The matrix $\PAmat$ exhibits
non-semisimple degeneracy at  $D=35$~nm, giving rise to an exceptional CPP
wave on Branch~II with  $q=(1.3484 + 0.2188 i)\ko$. For the same value of $D$,
an unexceptional CPP wave exists on Branch~I with  $q=(1.3484+0.0342 i)\ko$. 
Since $\Req$ is the same for both CPP waves, they have the same phase speed.
However, the exceptional CPP wave attenuates in the direction of propagation with a
higher rate than the unexceptional CPP wave.

No Dyakonov surface wave can be guided by the planar interface of materials
$\calA$ and $\calB$ when $\psi=23^\circ$. Therefore, Branch~I cannot be extended
to $D=0$.

\begin{figure}[!htb]
\centering
\includegraphics[width=4cm]{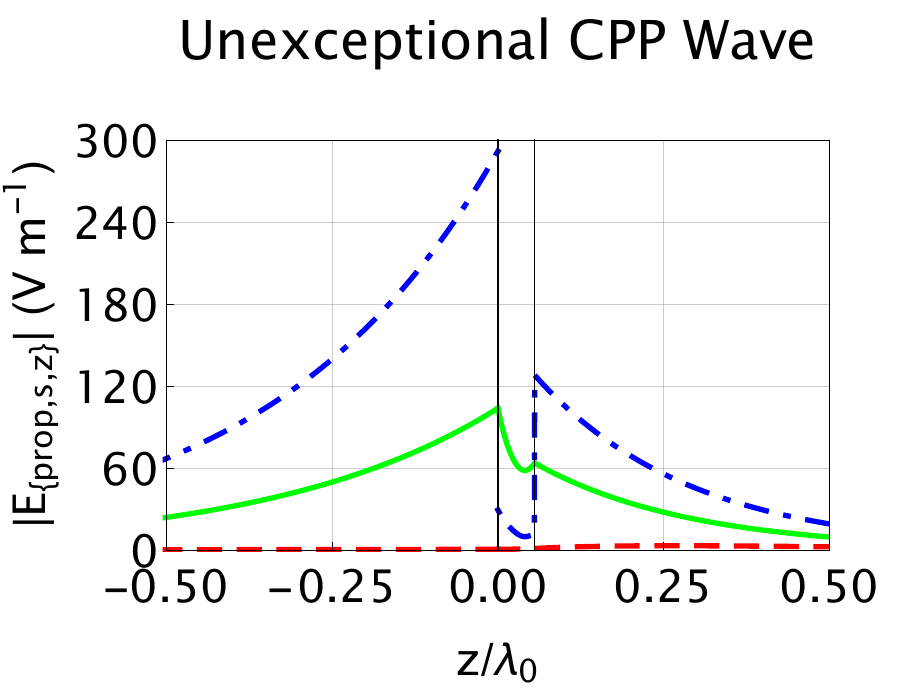} 
\includegraphics[width=4cm]{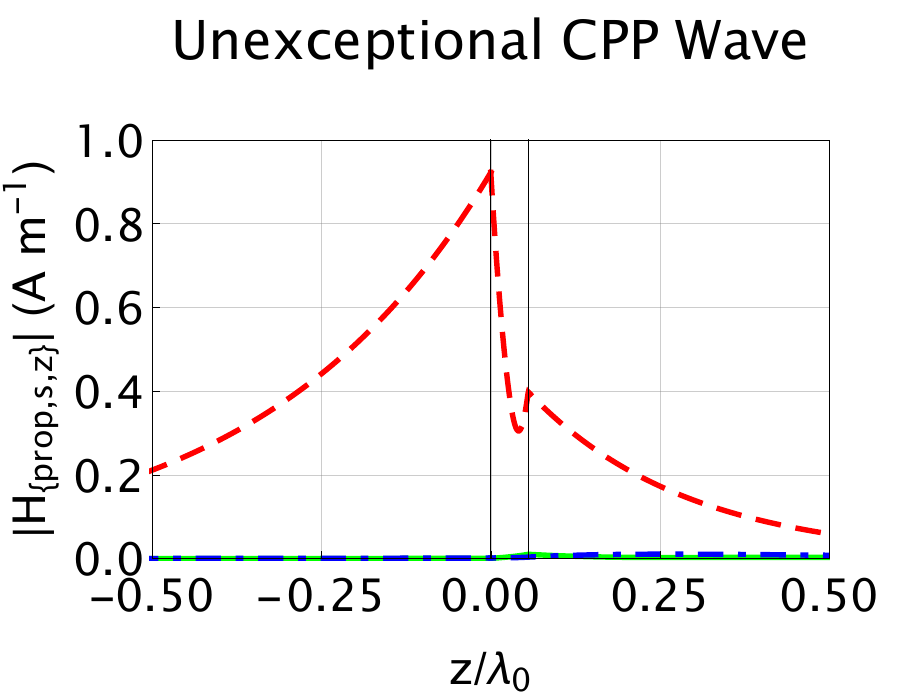}
\includegraphics[width=4cm]{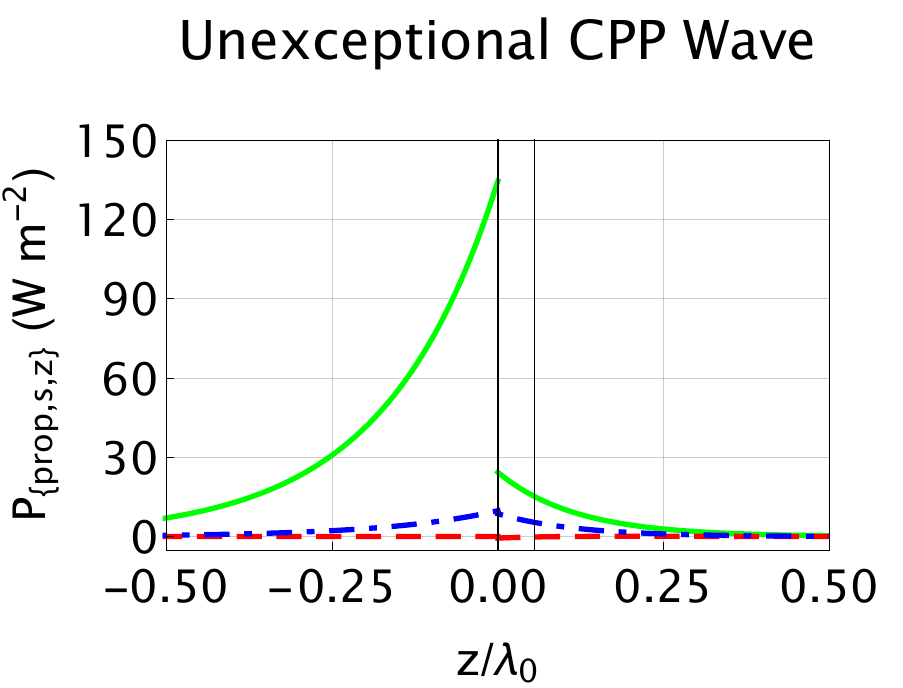} 
 \caption{\label{qFigure5}   
 $\vert\#E(z\uz)\. \#n\vert$ ,   $\vert\#H(z\uz)\. \#n\vert$ , and $\#P (z\uz) \. \#n$
 of the  unexceptional CPP wave [$q=(1.3484+0.0342 i)\ko$]   plotted versus $z/\lambdao$,
 when
 $\eps^s_\calA=1.5+0.5i$, $\eps^t_\calA=1.7896 + 0.4807 i$, $\epsB = -16.07 + 0.44 i$, $\eps_\calC=1.6066$, $\psi=23^\circ$, $D=35$~nm, and  $\lambdao=633$~nm. The normalization is 
such that  $\vert\#E(D\uz)\. \us\vert = 1$~V~m$^{-1}$. The left and right black vertical lines stand for the interfaces $\calB/\calC$ and $\calA/\calB$, respectively.
 Key:   $\#n = \uprop$ green solid curves; $\#n = \us$ red dashed  curves; $\#n = \uz$ blue broken-dashed curves.  
 }
\end{figure}

\begin{figure}[!htb]
\centering
 \includegraphics[width=4cm]{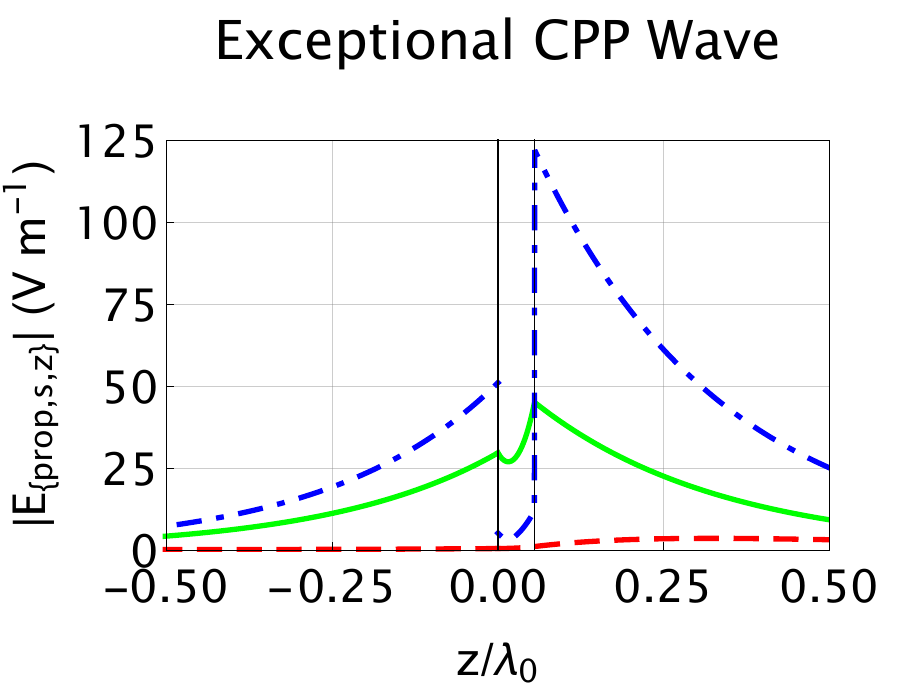}
 \includegraphics[width=4cm]{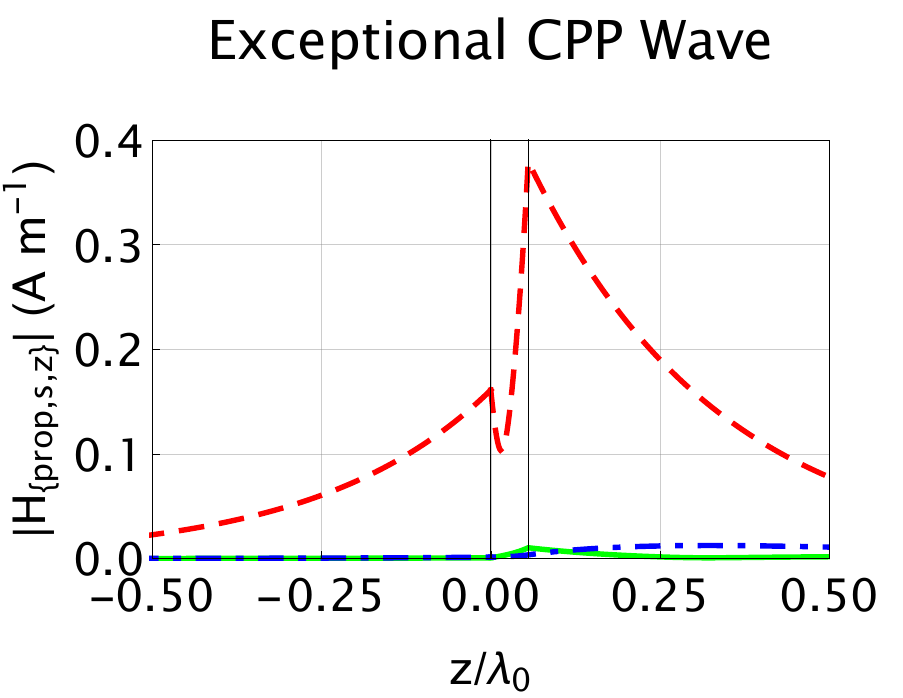}
 \includegraphics[width=4cm]{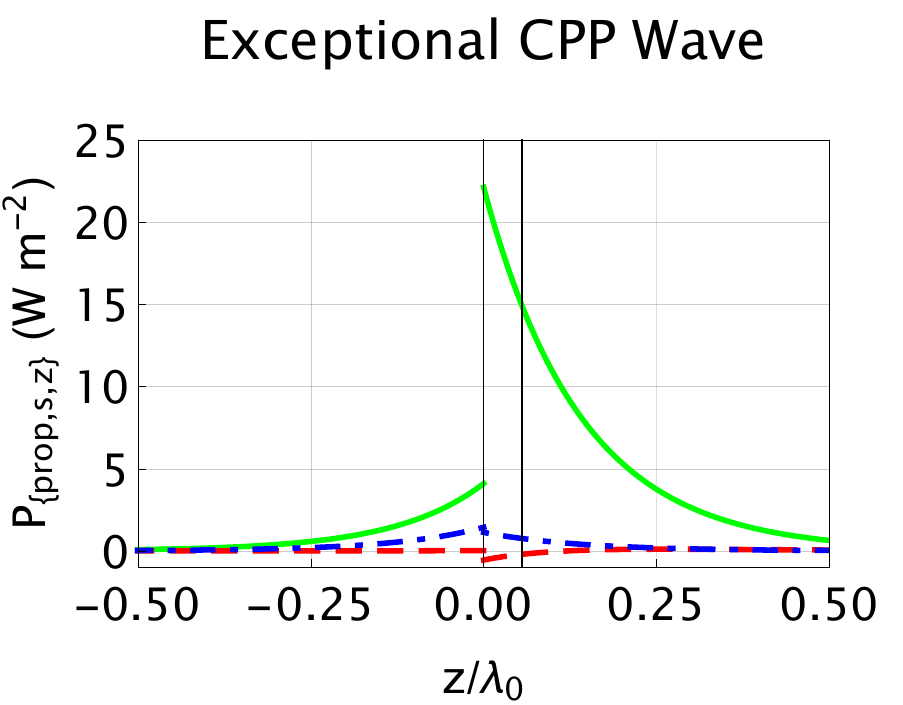}
 \caption{\label{qFigure6}   Same as Fig.~\ref{qFigure5} except
 for  the exceptional CPP wave [$q=(1.3484+0.2188 i)\ko$].
 }
\end{figure}

Given that the unexceptional and the exceptional CPP waves at $D=35$~nm
in Fig.~\ref{qFigure4} have the same phase speed, we decided to examine the
spatial profiles of $\#E(\#r)$ and $\#H(\#r)$, as well as
of the time-averaged Poynting vector
\begin{equation} 
\#P(\#r)=\displaystyle{\frac{1}{2}{\rm Re}\lec\#E(\#r)\times \blue{\#H^*(\#r)} \ric}
\end{equation}
of both waves. The magnitudes of components of all three quantities
 parallel
to the unit vectors $\uprop$, $\us=-\ux\sin\psi+\uy\cos\psi$, and $\uz$
evaluated for   $\#r=z\uz$ are plotted in Fig.~\ref{qFigure5} for
the  unexceptional CPP wave [$q=(1.3484+0.0342 i)\ko$], and
in Fig.~\ref{qFigure6} for
the  exceptional CPP wave [$q=(1.3484+0.2188 i)\ko$].

The spatial profiles
of the two CPP waves are very different from each other. The fields of the
unexceptional
CPP wave are higher in the isotropic material $\calC$ than in the
anisotropic material $\calA$ and, therefore, a much larger fraction of the energy of the
exceptional CPP wave is contained in material $\calC$ than in material
$\calA$. In contrast,  the fields of the
exceptional
CPP wave are higher in the anisotropic material $\calA$ than in the
isotropic material $\calC$ and, therefore, a much larger fraction of the energy of the
exceptional CPP wave is contained in material $\calA$ than in material
$\calC$. Furthermore, given that  $\vert\#E(D\uz)\. \us\vert = 1$~V~m$^{-1}$
in both figures, the maximum magnitudes of the Cartesian components 
of both fields and the time-averaged Poynting vector are higher for the
exceptional CPP wave than for the exceptional CPP wave.

\begin{figure}[!htb]
\centering
\includegraphics[width=4cm]{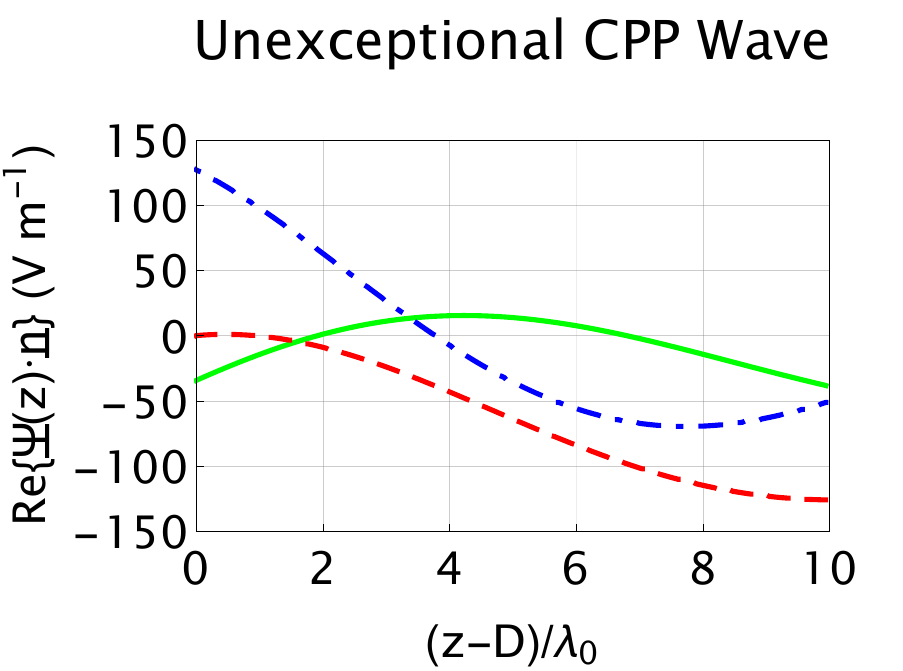} 
\includegraphics[width=4cm]{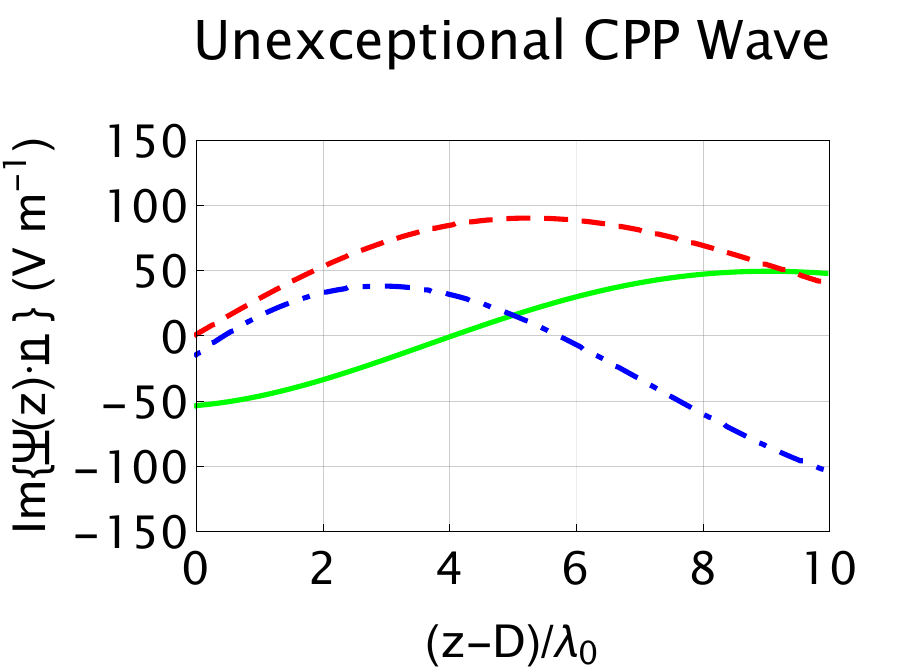}
 \caption{\label{qFigure7}   
 ${\rm Re}\lec  \#\Psi(z)\. \#n \ric$ and  ${\rm Im}\lec \#\Psi(z)\. \#n\ric$   of the  unexceptional CPP wave [$q=(1.3484+0.0342 i)\ko$]   plotted versus $(z-D)/\lambdao$,
 when
 $\eps^s_\calA=1.5+0.5i$, $\eps^t_\calA=1.7896 + 0.4807 i$, $\epsB = -16.07 + 0.44 i$, $\eps_\calC=1.6066$, $\psi=23^\circ$, $D=35$~nm, and  $\lambdao=633$~nm. The normalization is 
such that  $\vert\#E(D\uz)\. \us\vert = 1$~V~m$^{-1}$. 
 Key:   $\#n = \uprop$ green solid curves; $\#n = \us$ red dashed  curves; $\#n = \uz$ blue broken-dashed curves.  
  }
\end{figure}

\begin{figure}[!htb]
\centering
\includegraphics[width=4cm]{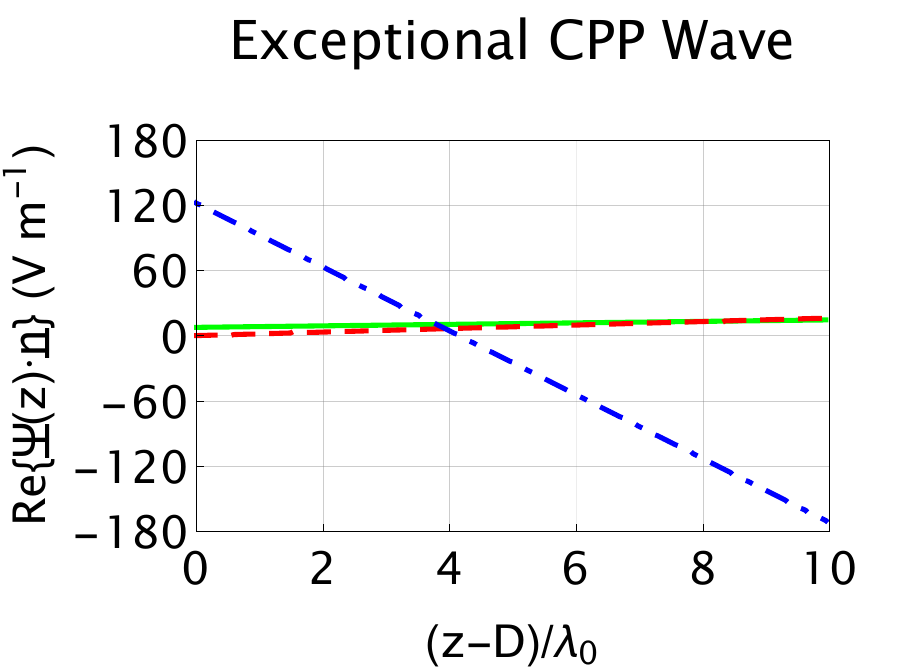} 
\includegraphics[width=4cm]{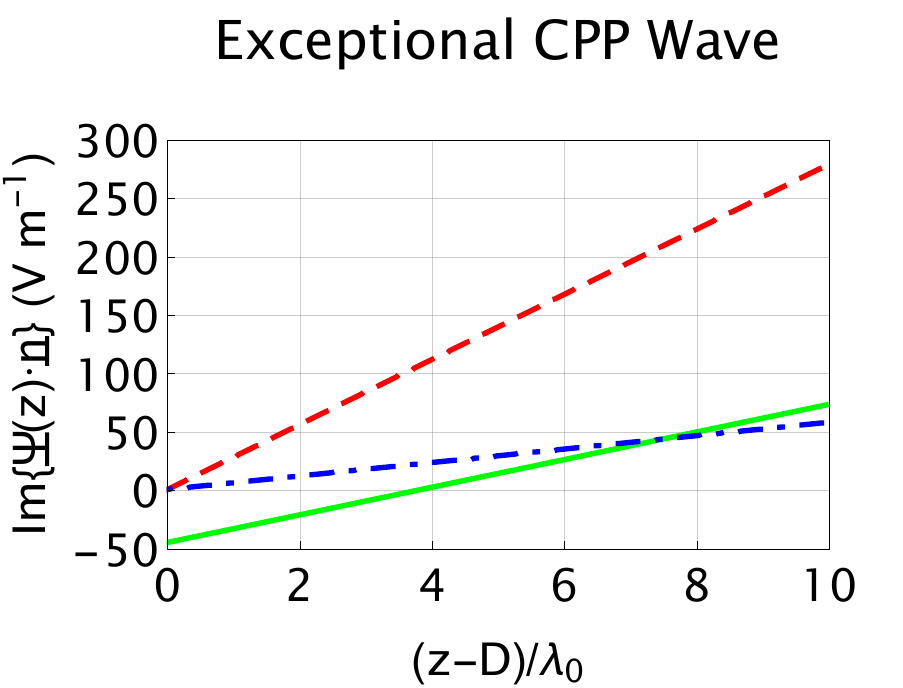}
 \caption{\label{qFigure8}   
 Same as Fig.~\ref{qFigure7} except
 for  the exceptional CPP wave [$q=(1.3484+0.2188 i)\ko$].
 }
\end{figure}

Figures~\ref{qFigure7} and \ref{qFigure8} provide a comparison of the spatial profiles
of the electric fields of both types of CPP waves in the anisotropic material
$\calA$. This comparison is warranted by the fact that $\PAmat$ is non-semisimply
degenerate for the exceptional CPP wave but not for the unexceptional CPP
wave, whereas $\PBmat$ and $\PCmat$ have the same eigenvalue characteristics
for both types of CPP waves. Since $\alpha_{\calA 1}\ne\alpha_{\calA 2}$ for the
unexceptional CPP wave, the
components of $\#\Psi(z)=\#E(z\uz) \exp\les-i\alpha_{\calA 1}(z-D)\ris$ vary with $z$ in an undulating fashion in Fig.~\ref{qFigure7}. On the other hand, $\alpha_{\calA 1}=\alpha_{\calA 2}$ for the
exceptional CPP wave, the
components of $\#\Psi(z)$ vary linearly with $z$
in Fig.~\ref{qFigure8}. Parenthetically, the increase in the magnitudes of the plotted quantities with $z$ in Fig.~\ref{qFigure8} should not cause alarm because attenuation
as $z\to\infty$ is due to $\exp \les - {\rm Im}\lec\alpha_{\calA 1}\ric(z-D)\ris$, but that has been
factored out of the definition of $\#\Psi(z)$.

\section{Closing remarks}\l{S4}
The objective of this paper is to introduce the concept of
exceptional compound plasmon-polariton waves. Such waves
are guided by a sufficiently thin metal film interposed between
two homogeneous dielectric materials of which at least one must
be anisotropic. Ordinarily, this arrangement will guide
unexceptional CPP waves, i.e., the fields on either side
of the metal film decay exponentially
with distance from the nearest metal/dielectric interface.
In contrast, the fields of an exceptional CPP wave decay on one
side of the metal film as the product of a linear function and
an exponential function of the   distance from the nearest metal/dielectric interface.

The simplest scenario for exceptional CPP-wave propagation was considered  in which one of the dielectric materials is uniaxial while the other is isotropic, and the metal is isotropic. Greater scope for  exceptional CPP waves is likely to be presented by scenarios in which 
 more than  one of the materials in the trimaterial structure is anisotropic. In particular, if the one (or more) of the materials 
 in a trimaterial structure  is biaxial, then the prospect of multiple exceptional CPP waves
 arises \c{Sci_Reports}. These are matters for future investigation.

\section*{Acknowledgements}
A. L. thanks the Charles Godfrey Binder Endowment at Penn State for ongoing support of his research. This work was supported in part by
EPSRC (grant number EP/S00033X/1).

\section*{Disclosures}

The authors declare that there are no conflicts of interest related to this article.

\end{document}